\documentclass[prl,twocolumn,showpacs,address]{revtex4}
\usepackage{amsfonts,amsmath,amssymb}
\usepackage{graphicx}

\begin{document}

\newlength{\figwidth}
\setlength{\figwidth}{0.45\textwidth}

\title{Tunnel ionization of open-shell atoms}

\author{Z. X. Zhao}
\author{T. Brabec}
\affiliation{Physics Department $\&$ Center for Research in Photonics, University of Ottawa, 150 Louis Pasteur,
Ottawa, ON K1N 6N5, Canada}
\begin{abstract}
A generalized ADK (Ammosov-Delone-Krainov) theory for ionization of open shell atoms is compared to ionization
experiments performed on the transition metal atoms V, Ni, Pd, Ta, and Nb. Our theory is found to be in good
agreement for V, Ni, Pd, and Ta, whereas conventional ADK theory overestimates ionization by orders of magnitude.
The key to understanding the observed ionization reduction is the angular momentum barrier. Our analysis shows
that the determination of the angular momentum barrier in open shell atoms is nontrivial. The Stark shift is
identified as the second dominant effect responsible for ionization suppression. Finally, these two effects cannot
explain the Nb data. An analysis of the electron spins suggests that Pauli blocking might be responsible for the
suppression of tunneling in Nb.
\end{abstract}
\maketitle

\noindent
Tunnel ionization theory was developed in several seminal works \cite{Smirnov66,Ammosov86}. All of these theories
are based on the single active electron (SAE) approximation, where only the weakest bound electron interacts with
the laser field. The most commonly used tunneling theories are ADK (Ammosov-Delone-Krainov) and MO-ADK
(molecular-ADK) theory \cite{Ammosov86}, which are in excellent agreement with experiments in noble gases and small
molecules \cite{Walker94}.

The situation is different for more complex systems. Theoretical predictions from ADK theory often overestimate
experimentally measured ionization yields by orders of magnitude \cite{Talebpour98,Smits04b}. Recently, a
quasi-analytical multi-electron theory of tunnel ionization was developed \cite{Brabec05}, revealing that the
SAE approximation loses its validity in complex materials. A comparison with ionization experiments in C$_{60}$
showed reasonable agreement. This theory opens the door to explore the new multi-electron physics associated with
tunnel ionization in complex systems.

In this paper we develop the theory of Ref. \cite{Brabec05} further to analyze tunnel ionization of the transition
metal atoms Vanadium (V), Nickel (Ni), Palladium (Pd), Tantalum (Ta), and Niobium (Nb). They are significantly
more complex than noble gas atoms due to an open $d$ shell and large angular and spin momenta. ADK theory
overestimates \cite{Smits04b} ionization in transition metal atoms by orders of magnitude. Our work reveals the
following new findings:

\noindent
(i) We find good agreement with V, Ni, Pd, and Ta experiments. This further corroborates the applicability of the
theoretical framework, developed here and in Ref. \cite{Brabec05}, to general complex systems.

\noindent
(ii) The first major effect responsible for the suppression of ionization is found to be the angular momentum
barrier. Due to the complexity of open shell atoms, determination of $l$ is not as
straightforward as in noble gases, but requires a careful analysis. This reveals unprecedented insight into the
structure and complexity of transition metal atoms. The second dominant contribution to the suppression of
tunneling comes from an increase of the ionization potential caused by the Stark shift. Our analysis 
demonstrates the importance of the Stark shift for tunnel ionization.

\noindent
(iii) The two effects discussed in (ii) cannot explain the strong ionization reduction found in Nb. An analysis
of the spins of s- and d-orbital electrons suggests that the reduction of tunnel ionization in Nb might be caused
by Pauli blocking, an effect central to many-body physics.

\noindent
(iv) Starting from the quasi-analytical multi-electron theory of of Ref. \cite{Brabec05}, a generalized
multi-electron ADK/MO-ADK theory is derived, which includes correction factors accounting for multi-electron
effects and angular momentum barrier.

Our analysis begins with the tunnel ionization rate in atomic units, as given in Ref. \cite{Brabec05},
\begin{equation}
w_m =  \frac{\Delta^{2(m+1)}}{(2\kappa)^{2|m|+1}|m|!} \left(\sum_l B_{lm} e^{-t} \right)^2 ,
\label{e1}
\end{equation}
where $l$ and $m$ refer to the angular momentum and magnetic quantum numbers of the field-free asymptotic
wavefunction of the weakest bound valence electron that is matched with the wavefunction of the tunneling
electron. In the case that $m$ is a non-degenerate, good quantum number, e.g. homonuclear diatomic molecules,
$w = w_m$, with $w$ the ionization rate. When $m$ is not a good quantum number, several $m$ contribute to the tunneling
wavefunction and $w = \sum_m w_m$. For $m$ degenerate, e.g. noble gas atoms, the ionization rate is determined
by $w = 1/(2l+1) \sum_m w_m$. Note that Eq. (\ref{e1}) corrects a misprint in Eq. (8) of Ref. \cite{Brabec05}
with respect to the position of $\sum_l$. Further, $B_{lm}=\sum_{m'}\mathcal{D}^{l}_{m'm}(\mathbf{R}) Q_{lm'}
C_{lm'}$, and $Q_{lm}$ is the normalization constant of the spherical harmonics \cite{Ammosov86,Brabec05}. The
rotation matrix ${\cal D}_{m'm}^l$ rotates the coordinate system by the Euler angle $\bf R$, and allows
calculation of ionization along an arbitrary direction, once the matching coefficients $C_{lm}$ have been
determined along one direction. Finally, $t = \int_{z_0}^{z_1} p_z dz$, $1/\Delta^2 = \int_{z_0}^{z_1} 1 / p_z dz$,
$p_z = \sqrt{ \kappa^2 - 2 z E + 2V(z) }$, and
\begin{equation}
V(z) = - {{\cal Z} \over z}  +  {\beta_{\small +} E \over z^2} - {\beta_+ \over 2z^4} + {l(l+1) \over 2z^2} .
\label{e7}
\end{equation}
The electric field $E$ is assumed to point along the $z$-direction, $\mathcal{Z}$ is the charge of the residual
ion after ionization, and $\beta = \beta_n$ and $\beta_+ = \beta_{n-1}$ refer to the polarizability of the
n-electron system before and after ionization \cite{Brabec05}. The Stark shifted \cite{Saenz02} and the field
free ionization potential of the weakest bound electron are given by $\kappa^2 /2 = I_p - (\beta - \beta_+) E^2/2$,
and $I_p$, respectively. The four terms in Eq. (\ref{e7}) represent the barrier coming from the Coulomb potential,
the laser polarization, the image charge, and the angular momentum, respectively. The last three terms are not
accounted for in ADK/MO-ADK theory. Integration is performed between the limits $z_0$ and $z_1$, where $z_1$ is
the outer turning point.

The lower integration limit $z_0$ is the point at which the laser dressed, tunneling wavefunction is matched to
the field free, asymptotic ground state, see Ref. \cite{Brabec05}. Matching has to be done at a field free
point. As complex systems are highly polarizable, the polarization field cancels the laser field at a point $a$,
where $\beta_{+} E / a^2 - a E \approx 0$. Solution of this equation gives $z_0 = a = \beta_{+}^{1/3}$. Due to
shielding the inner part of the system, $z\le a$, is approximately field free. In a metallic sphere, the matching point lies
on the surface.

When $V(z) \ll \kappa^2/2$ for $z_0 < z < z_1$ is fulfilled, the square root in $p_{z0}$ can be Taylor expanded.
In this limit, all the integrals can be solved analytically and contact can be made between Eq. (\ref{e1}) and
ADK/MO-ADK theory. The calculation is fairly involved; details will be given in a longer paper. The generalized
ADK/MO-ADK ionization rate is given by
\begin{eqnarray}
w_m &= &w_o \left( \frac{E}{2\kappa^2} \right)^{\alpha} \nonumber \\
\alpha & = &
{l(l+1)E \over \kappa^3} + {2\beta_+ E^2 \over \kappa^3} -{5\beta_+ E^3 \over 2\kappa^7 }
\label{e8}
\end{eqnarray}
where $w_o$ refers to the conventional ADK/MO-ADK ionization rate,
\begin{equation}
w_o = \frac{|\hat{C}_{lm}Q_{lm}|^2} {2^{|m|}|m|!\kappa^{2{\cal Z}/\kappa-1}}
\left(\frac{2\kappa^3}{E}\right)^{2{\cal Z} / \kappa -|m|-1} \exp \left( {-2 \kappa^3 \over 3E} \right).
\label{e9}
\end{equation}
The ADK/MO-ADK matching coefficient in Eq. (\ref{e9}) is related to the matching coefficient in Eq. (\ref{e1})
by $\hat{C}_{lm} = C_{lm}e^{\kappa z_0}z_0^{-{\cal Z}/\kappa}$. We use here $w = 1/(2l+1) \sum_m w_m$. The three
correction terms in $\alpha$ come from the angular momentum barrier, the laser polarization barrier, and the
image charge term, respectively. The l-barrier is not contained in the ADK/MO-ADK theory. The $l$-dependence in
ADK/MO-ADK theory comes from the spherical harmonics and the matching coefficient. In the limit of $l=0$ and small
polarizability, Eq. (\ref{e8}) reduces to the conventional ADK/MO-ADK ionization rate, $w_m = w_o$. A comparison
of Eq. (\ref{e8}) with Eq. (\ref{e1}) shows excellent agreement for the transition metal atoms investigated here.
Therefore, we use Eqs. (\ref{e8}) - (\ref{e9}) throughout the paper. The applicability of (\ref{e8}) for complex
molecules will be investigated in a future work.

\begin{figure}
\includegraphics*[width=3.3in]{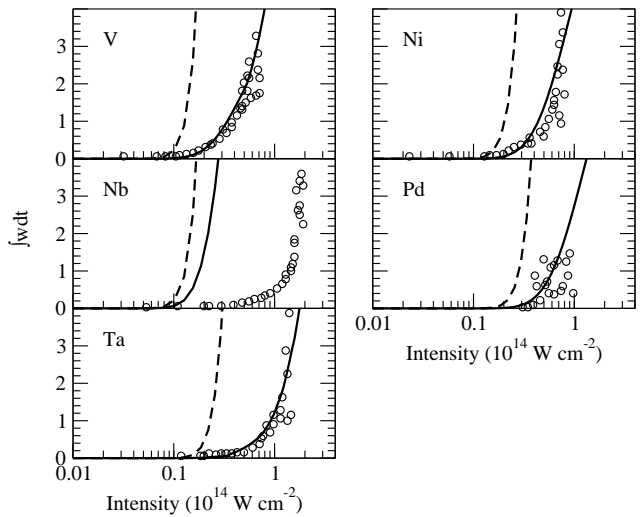}
\caption{\label{f1} Tunnel ionization yield $\int w dt$ in V, Ni, Nb, Pd, and Ta versus laser
peak intensity. The open circles denote the experimental results from Ref. \cite{Smits04b}. The dashed and
full lines refer to the theoretical results obtained from ADK theory, and from our theory, Eq. (\ref{e8}),
respectively. Gaussian laser pulse parameters: wavelength $1.5 \mu$m and FWHM width $50fs$.}
\end{figure}

Key to understanding tunnel ionization in transition metal atoms is knowledge of the angular momentum $l$ of the
asymptotic wavefunction of the weakest bound valence electron. Quantum chemistry codes have been used to determine
$l$ for small molecules and noble gas atoms \cite{Ammosov86,Kjeldsen05a}. As Hartree-Fock (HF) is sufficient to model
these systems, the asymptotic wavefunction can be obtained from the highest occupied molecular orbital. We find
that this method is also applicable to Pd and gives $l=2$, see the discussion below. This allows us also to extract
$\hat{C}_{lm}$ for Pd from the numerical analysis and to compare it with the ADK matching coefficient
\cite{Ammosov86}, as determined by quantum defect theory. The ADK matching coefficient is commonly used to calculate
tunnel ionization of atoms. The numerical result is found to be by a factor of 1.2 smaller. This difference is
insignificant and corroborates the validity of quantum defect theory. Hence, the ADK matching coefficient is used
for all the other transition metal atoms, where a numerical determination is not possible.

In all other transition metals the wavefunction undergoes a complex configuration mixing, when the valence electron
is pulled away from the nucleus to the asymptotic region. Quantum chemistry codes are not sensitive to the
asymptotic part of the wavefunction, as they optimize the binding energy, which mainly depends on the wavefunction
close to the nucleus. Although an accurate description of the asymptotic wavefunction is in principle possible
within a configuration interaction (CI) analysis, in practice it requires an unreasonably large number of
configurations. Therefore, due to the lack of proper numerical methods $l$ is inferred from the configuration data
given in table \ref{t1}.

Our analysis of the transition metal atoms starts with Vanadium. Whereas in noble gases $l$ is determined by the
highest occupied orbital, this is no longer possible here. Although the highest occupied orbital of V is a 4s
orbital, the following more careful analysis of the total angular momentum $L$ shows that $l=0$ is forbidden. The
total angular momentum of the atom and ion ground state is $L=3$ and $L_+=2$, respectively, see table \ref{t1}.
The plus subscript refers to ions and ion-parameters throughout the paper. From the angular momenta $L$ and $L_+$
a range of possible values of $l$ can be derived. This is done by using the field free, asymptotic, atomic ground
state, which can be approximated as the tensor product of the ionic ground state and of the valence electron far
away from the nucleus. As a result, the total angular momentum of the ion plus valence electron, as determined
by the angular momentum addition theorem, has to be equal to the total angular momentum of the neutral atom. This
is only possible when $l$ fulfills the condition $L_+ - l \le L \le L_+ + l$, yielding $1 \le l \le 5$. The
possible range of $l$ can be further narrowed down by utilizing the fact that parity also has to be conserved. As
V and V$_+$ have even parity, the parity of the asymptotic valence electron wavefunction must be even, too. This
requires $l$ to be even and thus, $l=2,4$. As the 3d orbitals are energetically close to the highest occupied 4s
orbitals, the dominant tunneling channel comes from a 3d-orbital with $l=2$.

We would like to note three points associated with the above analysis. First, the conservation laws were applied
to the field free, asymptotic ground state, which is the initial state for tunneling. Angular momentum and parity
are of course not conserved during tunnel ionization. Second, the atomic ground state close to the nucleus and in
the
asymptotic limit is described by different configurations. This mixing of configurations in the time independent
ground state should not be confused with a time-dependent reconfiguration. Third, we assume that the atom relaxes
adiabatically during ionization into its ionic ground state. When an excited state is populated after ionization,
which is termed shake-up \cite{Litvinyuk05}, the ionization potential increases by the energy difference between
ion excited and ground state, $\Delta I$. In transition metal atoms $\Delta I \approx 0.3-1$eV is small, so that
the shake-up channel might be important. Whereas it would be surprising when tunnel ionization leaves ions
dominantly in an excited state, due to the complexity of the problem it also cannot be rigorously ruled out. Our
angular momentum analysis gives $l=0,2$ for the shake-up channel. In case $l=2$, the shake up channel gives
a small correction to the fundamental channel and our results remain unchanged. In case $l=0$, our theory would
predict a too large ionization yield, close to ADK theory. The $l=0$ shake-up channel is excluded based on the fact
that $l=2$ explains the experimentally observed ionization suppression very well. We suggest an experiment to
corroborate our conclusion. The first excited state in transition metal atoms is non-dipole allowed and therewith
metastable. Therefore, the population of the ion states after ionization can be probed by exciting dipole allowed
transitions and measuring fluorescence or the loss experienced by the probing laser. For example, in V$_+$ the
lowest dipole allowed transitions from the $3d^4$ ground state and the first excited (metastable) $3d^34s$ state
can be probed, with transition energies of $\approx 4.5$eV and $4$eV, respectively. In dependence on whether the
shake-up channel has $l=0$ or $2$, one would find a predominant population of the $3d^34s$ state or
of the $3d^4$ ground state, respectively.

Given the simple approach used here and experimental uncertainties, our theory is in good agreement with the
experimental data of Ref. \cite{Smits04b}. In Fig. \ref{f1} the ionization yield $\int w[E(t)] dt$ is plotted
versus the laser peak intensity. For pulse parameters and shape of the laser electric field $E(t)$ see the
figure caption. Our theory (full line) explains the difference between experiment (empty circles) and ADK
theory (dashed line). Half of the difference is due to the $l$-barrier which is not accounted for in ADK/MO-ADK theory. As $\beta_+$ is small, the role of the laser polarization induced
barrier is insignificant. The other half of the difference is accounted for by the Stark shift, which reduces
ionization by increasing the ionization potential. The large Stark effect arises from the big difference in
polarizabilities $\beta$ and $\beta_+$, see table \ref{t1}. As the Stark effect increases with laser intensity,
its effect is most pronounced at high intensities. Therefore, the ionization yield rises gradually and
reproduces the shape of the experimental curve, in contrast to the abrupt rise predicted by ADK/MO-ADK theory.
Switching off the l-barrier shows that the Stark effect requires the l-barrier to suppress ionization
effectively. Without the l-barrier ionization would be shifted to lower intensities, where the role of the
Stark effect is insignificant.

Part of the large discrepancy between ADK/MO-ADK theory and experiment might arise from the fact that ADK/MO-ADK
overestimates tunnel ionization in the above barrier limit. The Coulomb barrier is suppressed at an electric
field $E_{bs} = I_p^2/4$, which corresponds to a laser intensity of $7 \times 10^{12}$W/cm$^2$ in V. Hence,
barrier suppression in V occurs before ADK/MO-ADK theory shows that significant ionization takes place. In H and He,
ADK/MO-ADK theory overestimates above the barrier ionization rates by a factor of 2-4 \cite{tong05}, which is by
far not enough to explain the difference in Fig. \ref{f1}. In transition metals the magnitude of the overestimation
is unknown. Nevertheless, we believe that barrier suppression is not the dominant factor responsible for the
difference between ADK/MO-ADK theory and experiment. Significant ionization usually takes place  around the intensity at
which the barrier is suppressed. In V ionization becomes dominant in the range between $4 \times 10^{13}$W/cm$^2$
and $7 \times 10^{13}$W/cm$^2$, which is by a factor of 5-10 larger than the barrier suppression intensity. It
appears to be counterintuitive that ionization should only start at intensities so much higher than the barrier
suppression intensity. In our opinion the large difference points to the fact that ADK/MO-ADK theory, which is
exclusively based on the Coulomb barrier, underestimates the true tunneling barrier. This is corroborated by our
theory, Eq. (\ref{e7}). Due to $l$-barrier and Stark shift, the tunneling barrier is not suppressed over the laser
intensity range used in Fig. \ref{f1}.

For Ni and Pd we get similarly good agreements as for V, see Fig. \ref{f1}. For Ni our analysis yields $l=2$.
In Pd determination of $l$ is straightforward. The total angular momentum changes from $L=0$ to $L_+ = 2$.
The configuration changes from 4d$^{10}$ to 4d$^{9}$, showing that the tunneling electron comes from a d-orbital
and that $l=2$, in agreement with the HF analysis discussed above.

In Ta it was not possible to unambiguously identify $l$. From table \ref{t1} we find two relevant ionization
channels. The possible angular momentum values are $l = 0,2,4,6$. Hence, ionization from an s-orbital and from
a d-orbital are possible. In V tunneling from a d-orbital is preferred, probably because breaking of the closed
subshell 4s$^2$ configuration is energetically unfavorable. We suspect that tunneling from a d-orbital in Ta
might be favored for similar reasons. This is supported by comparing theory and experiment for $l=2$, which
yields good agreement. For $l=0$ the experimental data cannot be explained.

Finally, for Nb we could not get agreement with experimental data. The possible values of $l$ are $l=0,2,4$.
Niobium does not have a closed s-subshell like V, Ni, and Ta, which might favor tunneling of a d-electron.
Therefore, we use $l=0$ in Fig. \ref{f1}. We have also performed test calculations for $l=2$ and $4$. These
also do not give satisfactory agreement, corroborating our conclusion that the $l$-barrier is not
responsible for the observed ionization suppression. Further, ionization due to dimers and higher clusters
was excluded experimentally \cite{albert1}.

A spin analysis of the s- and d-electrons shows that the spins of the tunneling and bound electrons are all
parallel in Nb. In all other transition metal atoms the spins of the bound electrons are parallel, however
the tunneling spin points in the opposite direction. Therefore, in V, Ni, Pa, and Ta, the exchange effect
between tunneling and bound electrons is zero. In contrast to that, the tunneling electron in Nb feels the
exchange effect with the bound electrons. The exchange effect makes it harder for the tunneling electron to
pass the other open shell electrons and to escape. In the language of many-body physics this is termed Pauli
blocking.

Based on the fact that none of the other mechanisms can explain the Nb experiments, we speculate that Pauli
blocking is responsible for the observed ionization suppression. A quantitative proof will require a generalization
of the current ionization theory to account for the exchange effect, which will be subject to future work. 


\begin{table}[htp]
\begin{tabular}{*{2}{l}|*{1}{l}*{1}{c}*{1}{l}|*{1}{l}*{1}{c}*{1}{l}|*{2}{c}}
\hline\hline
&\multicolumn{1}{c}{Z}&\multicolumn{3}{c}{Neutral}&\multicolumn{3}{c}{Ion} &Ip (eV)&{$l$}     \\

\cline{1-10}
&&Conf.&$^{2S+1}L_{\!_J}$ & $\beta$  &Conf.&$^{2S+1}L_{\!_J}$   &$\beta_+$&   & \\

V   &23& 3d$^3$4s$^2$ &$^4$F$_{\frac{3}{2}}$  &12.4 &3d$^4$ & $^5$D$_0$ &3.1
 & 6.75
&2 \\
Ni &28 &3d$^8$4s$^2$ &$^3$F$_4$  &6.8    &3d$^9$ &$^2$D$_{\frac{5}{2}}$ &1.4
 & 7.64
& 2\\
Pd &46 &4d$^{10}$ &$^1$S$_0$ &4.8 &4d$^{9}$ &$^2$D$_{\frac{5}{2}}$ &1.8
&8.34
&2\\
Ta  &73 &5d$^3$6s$^2$ &$^4$F$_{\frac{3}{2}}$ &13.1 &5d$^3$6s &$^5$F$_1$& 7.7
&7.89
&2 \\
Nb &41& 4d$^4$5s &$^6$D$_{\frac{1}{2}}$   &15.7  &4d$^4$ &$^5$D$_0$  &4.7
&6.76
&0 \\
\hline
\hline
\end{tabular}
\caption{\label{t1}Table for the transition metal elements  V, Ni, Pd, Ta and Nb, containing the total charge of
the nucleus, configuration, total spin and angular momentum, polarizability ($\AA^3$) of neutral and singly charged
ion, ionization potential $I_p$ and the angular momentum $l$ of the tunneling electron. The plus subscript refers
to singly charged ions and their properties throughout the paper. The polarizability is calculated by using density
functional theory in the local density approximation \cite{Zangwill84};  $\beta$ agrees with published data, see http://
webbook.nist.gov;  for $\beta_+$ no data is available for comparison.
The configuration data is taken from http://physics.nist.gov/PhysRefData/Handbook/index.html. }
\end{table}


\end{document}